%% file: main0729F.tex
\documentclass{article}
\usepackage[margin=1in]{geometry}
\usepackage{amsmath,amssymb,amsfonts}
\usepackage{booktabs}
\usepackage{graphicx}
\usepackage{natbib}
\usepackage{hyperref}
\usepackage{caption}
\usepackage{subcaption}
\usepackage{float}
\usepackage{placeins}
\usepackage{array}
\usepackage{microtype}
\hypersetup{
  colorlinks=true,
  linkcolor=blue,
  citecolor=blue,
  urlcolor=blue,
  pdftitle={Robust Hedging Valuation Adjustment for Deep Hedging Policies under Market Frictions},
  pdfauthor={Takayuki Sakuma},
  pdfkeywords={deep hedging, hedging valuation adjustment, robust risk measurement, transaction costs, no-transaction band}
}
\newcommand{\E}{\mathbb{E}}
\newcommand{\KL}{D_{\mathrm{KL}}}
\newcommand{\RHVA}{\mathrm{RHVA}}
\newcommand{\RFA}{\mathrm{RFA}}
\newcommand{\RMA}{\mathrm{RMA}}
\newcommand{\CVaR}{\mathrm{CVaR}}
\newcommand{\RAdj}{\mathrm{RAdj}}

\title{Robust Hedging Valuation Adjustment for Deep Hedging Policies under Market Frictions}
\author{Takayuki Sakuma\footnote{e-mail: tsakuma@soka.ac.jp.}\\
Faculty of Economics and Business Administration, Soka University}
\date{\today}

\begin{document}
\maketitle

\begin{abstract}
Hedging a derivative position under transaction costs and market frictions requires a trading rule that adapts to changing conditions. Deep hedging trains a neural policy for this task but policy training does not determine whether a trading desk can afford to run the policy. We apply robust hedging valuation adjustment (HVA) as a post-training valuation-adjustment layer that evaluates tracking-loss CVaR together with explicit funding and margin add-ons. The funding and margin add-ons share the same KL uncertainty set as HVA. For each policy, a single common-stress tilt computes HVA, funding and margin jointly and a trading desk can get one internally consistent reserve instead of the three separately. We compare classical hedge policies with learned hedge specifications across three market environments with different liquidity. No single specification dominates in every market. Under the strict tracking-risk budget, gamma-wide classical bands are selected in High and Middle Liquidity while sparse learned execution is selected in Low Liquidity. At looser validation budgets wider classical bands are generally selected.
\end{abstract}


\section{Introduction}

A hedge that works well in one market condition can be too expensive or too risky in another because transaction costs, market impact, and liquidity vary with the trading environment. Instead of a closed-form rule such as a Black--Scholes delta hedge \citep{BlackScholes1973} or a Whalley--Wilmott no-transaction band \citep{WhalleyWilmott1997}, deep hedging trains a policy on cost and risk objectives over simulated market paths \citep{Buehler2019DeepHedging}. A policy can reach a low training loss while trading too often and running up spread and impact costs. It can also trade too little and leave a large terminal hedge error.

The literature on hedging under transaction costs begins with three classical results. \citet{Leland1985} shows that proportional transaction costs change replication by altering the effective volatility. \citet{WhalleyWilmott1997} derive a no-transaction band around the Black--Scholes hedge. If the current position is within that band, rebalancing to close the small remaining gap would trim hedge-error variance but not enough to offset the transaction cost of trading. \citet{AlmgrenChriss2001} develop an optimal-execution framework in which trading cost grows with the square of trade size.

Jump risk also complicates hedge evaluation because a local delta hedge cannot eliminate it. \citet{Merton1976} introduces discontinuous stock returns as the source of this risk. \citet{KennedyForsythVetzal2009} study dynamic hedging under jump diffusion with transaction costs and choose trades by balancing risk reduction against transaction-cost penalties. \citet{Sepp2012} analyzes delta-hedging error distributions under discrete trading and transaction costs in diffusion and jump-diffusion models. 

\citet{Buehler2019DeepHedging} formulate the deep-hedging training problem and treat transaction costs, liquidity constraints, and risk limits as a neural stochastic-control problem. \citet{Murray2022DeepHedgingRL} extend the approach with actor-critic learning across portfolios and risk aversions. \citet{BuehlerMurrayWood2022DeepBellman} develop a Bellman version for arbitrary portfolios across market states. \citet{Imaki2021PFHedge} impose an interpretable no-transaction-band architecture whose output is executed by nearest-boundary projection and \citet{ArzelLehdili2026} add structural stochastic-control priors to learned bands. These methods enlarge or constrain the policy class but do not by themselves determine whether a learned rule is preferable once reserve cost and residual risk are measured.

Other deep-hedging studies change the training methods. \citet{CaoHullPoulos2021} use reinforcement learning for derivatives hedging under transaction costs and \citet{HorvathTeichmannZuric2021} study deep hedging under rough volatility. \citet{Mueller2024FastDeepHedging} develop faster optimization for deep hedging and illustrate it on a path-dependent option. \citet{Godin2016CVaR} minimizes CVaR in a global dynamic-hedging problem with transaction costs. \citet{Lutkebohmert2022Robust} make deep-hedging training robust to parameter uncertainty. \citet{Neagu2024MarketImpact} embed convex and persistent market impact in a deep reinforcement-learning policy. \citet{HuangLawryshyn2025} evaluate reinforcement-learning hedgers under market impact, slippage, and transaction costs. \citet{Shinozaki2024} reviews deep hedging and deep calibration as practical deep-learning tools for finance. 

A learned hedge's position can also function as a statistical-arbitrage trade. \citet{HorikawaNakagawa2024} relate the difference between deep and delta hedging to statistical arbitrage in a complete-market setting. \citet{Francois2025StatArb} show in an incomplete-market experiment that the speculative position depends on how strongly the risk measure penalizes adverse outcomes. That risk is why we impose the explicit tracking-CVaR budget in this paper and a lower reserve counts as better hedging only if residual risk stays within that same constraint. \citet{GongVentreOHara2021} use neural networks to trace an efficient hedging frontier. We add a valuation-adjustment layer for trained policies. It decomposes a policy-dependent reserve into transaction-cost HVA, a simplified funding add-on, and a simplified margin add-on, and evaluates their sum under one policy-specific KL stress. 

The valuation-adjustment layer also needs to calculate funding and margin because the hedge path itself creates a cash balance and short-horizon hedged losses. \citet{Piterbarg2010} shows that collateral agreements and funding costs affect derivative valuation and \citet{PallaviciniPeriniBrigo2011} build a collateralized counterparty valuation framework with funding, collateral, netting, and rehypothecation. We use this idea at the policy level and the hedge path generates a funding cash deficit and a funding add-on. \citet{GreenKenyon2015MVA} define MVA as the funding cost of initial margin and emphasize the expected margin profile. Similarly we apply an expected-shortfall margin rule from short-horizon hedged losses and compute a margin add-on. \citet{BurgardKjaer2011} contribute to the broader semi-replication xVA literature that jointly treats counterparty risk and funding. Our reserve concept comes from the hedging valuation adjustment (HVA) literature, and \citet{Burnett2021HVA} formulates HVA as the value effect of transaction costs. \citet{Sakuma2026RobustHVA} develops robust HVA under liquidity-demand stress. Relative to that paper, which compares classical no-transaction bands under fixed-radius and fixed benchmark-stress conventions, the present paper adds trained neural policies, policy-generated funding and margin costs, common-stress aggregation, and validation/test selection.
  
This paper treats deep hedging as a policy generator and the robust reserve layer as the policy evaluator. In the present simulation framework, the non-robust transaction-cost HVA and the simplified funding and margin add-ons are expected path costs under the reference model, but their values depend on model and parameter choices. The robust layer instead evaluates the total reserve over a disclosed relative-entropy neighborhood. The fixed-radius convention supplies a common entropy budget across policies, although it does not preserve the benchmark-stress label across policy-specific cost distributions. The contribution is threefold. First, it replaces the single classical band with a policy set that includes deep hedging, a learned no-transaction band, and sparse-execution rules. Second, it adds robust funding and margin add-ons under the same KL stress as robust HVA, with the total adjustment evaluated under one common-stress path distribution. Third, it ranks policies under a tracking-risk budget. A policy that skips the hedge almost entirely can look best by cost alone but it leaves the desk exposed to large unhedged risk. 

\section{Framework}
\label{sec:framework}

\subsection{Price and liquidity processes}
\label{sec:model}

We consider the risk-neutral Merton jump-diffusion model
\begin{equation}
    \frac{dS_t}{S_{t-}}=(r-\lambda_J\kappa_J)dt+\sigma dW_t+(J-1)dN_t,
    \qquad
    \log J\sim N(\mu_J,\sigma_J^2),
    \label{eq:merton-process}
\end{equation}
where $N_t$ is a Poisson process with intensity $\lambda_J$ and
\begin{equation}
    \kappa_J=\E[J-1]=\exp(\mu_J+\tfrac12\sigma_J^2)-1.
\end{equation}
We use the analytic Merton price and delta \citep{Merton1976} and they reduce to the Black--Scholes quantities \citep{BlackScholes1973} if $\lambda_J=0$.

We simulate the model on an equally spaced training or evaluation grid and, at each step, approximate jump occurrence by
\begin{equation}
    \Delta N_i=\mathbf 1\{U_i<\lambda_J\Delta t\}.
    \label{eq:bernoulli-jump}
\end{equation}
The stock is updated as
\begin{equation}
    \log\!\left(\frac{S_{i+1}}{S_i}\right)
    =\left(r-\lambda_J\kappa_J-\tfrac12\sigma^2\right)\Delta t
    +\sigma\sqrt{\Delta t}\,Z_i+\Delta N_iY_i,
    \qquad Y_i\sim N(\mu_J,\sigma_J^2),
    \label{eq:discrete-merton}
\end{equation}
where $U_i\sim U(0,1)$ and $Z_i\sim N(0,1)$, so this Bernoulli discretization permits at most one jump per step. Liquidity follows a two-state Markov chain with normal and stressed multipliers $M_i\in\{1,m_s\}$ with transition matrix
\begin{equation}
    P_M=
    \begin{pmatrix}
    p_{NN} & 1-p_{NN}\\
    1-p_{SS} & p_{SS}
    \end{pmatrix}.
    \label{eq:liquidity-transition}
\end{equation}
This starts from its stationary distribution with stressed-state probability
\begin{equation}
    \pi_S=\frac{1-p_{NN}}{2-p_{NN}-p_{SS}}
\end{equation}
and is simulated independently of the Brownian and jumps. The neural policies do not observe $M_i$ and the multiplier is applied only in post-training evaluation. 

\subsection{Trained deep-hedging policies}
\label{sec:deephedge}

Deep hedging is a trained stochastic-control method. A market model generates price paths and at each date a neural network observes the current state and outputs a hedge position \citep{Buehler2019DeepHedging}. Stochastic gradient descent can train the policy by backpropagating through the trading path. At date $t_i$, a neural policy observes
\begin{equation}
    Z_i=(\log(S_i/K),\sqrt{\tau_i},\tau_i,q_{i-1}),
\end{equation}
where $\tau_i=T-t_i$ and $q_{i-1}$ is the executed position carried into the date. A feedforward network proposes
\begin{equation}
    \widetilde q_i^{\theta}=\Pi_{\theta}(Z_i)
\end{equation}
and the policy executes $q_i^{\theta}=\widetilde q_i^{\theta}$.

Let $D_i=e^{-rt_i}$ and $\widetilde S_i=D_iS_i$. The discounted hedging gain and terminal hedge loss are
\begin{align}
    G_T^{\theta}&=\sum_{i=0}^{N-1}q_i^{\theta}(\widetilde S_{i+1}-\widetilde S_i),\\
    L_T^{\Pi_{\theta}}&=D_N(S_N-K)^+-G_T^{\theta}
    \label{eq:terminal-loss}
\end{align}
and the neural networks minimize
\begin{equation}
    \mathcal L(\theta)=\CVaR_{95}(L_T^{\Pi_\theta})
    +\lambda_s\overline C_s^{\Pi_\theta}
    +\lambda_{\kappa}\overline C_{\kappa}^{\Pi_\theta}
    +\lambda_X\overline X^{\Pi_\theta}
    +\lambda_N\overline N^{\Pi_\theta},
    \label{eq:training-objective}
\end{equation}
where the last four terms are batch-average spread cost, quadratic impact, dollar turnover, and a differentiable trade-count penalty. For a batch of size $B$, the implementation estimates $\CVaR_{95}$ by averaging the largest $\lceil0.05B\rceil$ losses. The soft trade count replaces each trade indicator by $\operatorname{sigmoid}((|\Delta q_i|-0.02)/0.01)$ and includes an analogous terminal-unwind term. The CVaR representation follows \citet{RockafellarUryasev2000}. This objective adds explicit penalties on cost and turnover to the terminal-loss CVaR because $L_T^{\Pi_\theta}$ carries no transaction costs, and different trading patterns that reach the same terminal loss can still differ sharply in turnover or trade count. 

\subsection{Classical and learned execution rules}
\label{sec:policies}

This subsection introduces the no-transaction-band rules (Black--Scholes, Merton, and gamma-wide), the learned no-transaction-band (NTB) policy, and a post-training sparse-execution filter applied to the neural hedger defined above.

All policies begin from $q_{-1}=0$ and let $d_i$ be a target delta and $b_i$ a half-width. The no-transaction-band rule is
\begin{equation}
    q_i
    =\min\{\max(q_{i-1},d_i-b_i),d_i+b_i\},
    \qquad i=0,\ldots,N-1.
    \label{eq:projected-band}
\end{equation}
The opening position is the point in the initial band nearest to zero, and a later breach moves the hedge only to the nearest boundary. The Black--Scholes fixed-width rule sets $d_i=\Delta_i^{BS}$, while the separate Merton fixed-width family uses $d_i=\Delta_i^M$ in the jump environments. The gamma-wide family below is centered on $\Delta_i^{BS}$ in all three environments. For both fixed-width rules, the base width $b_0$ is selected from a grid of candidate values, given in Section~\ref{sec:design}.

For the gamma-wide rule, we adapt the gamma-aware band construction of \citet{Sakuma2026RobustHVA}. To widen the no-transaction band in states where the target delta is expected to move quickly, we set
\begin{equation}
    b_i=\min\{3b_0,b_0(1+\widetilde G_i)\},
    \qquad
    \widetilde G_i=\min\!\left\{
    \frac{|\Gamma_i|S_i\sigma\sqrt{\Delta t}}{Q_{90}},5
    \right\},
    \label{eq:gamma-wide}
\end{equation}
with $b_0$ selected from the same candidate grid. $\Gamma_i$ is the Black--Scholes gamma of the option at date $t_i$. The scalar $Q_{90}$ is the 90th percentile of gamma intensity on an independent 20,000-path sample and is fixed across validation and test seeds.

Third, we follow \citet{Imaki2021PFHedge} for the learned no-transaction-band policy in which the network outputs a band instead of a hedge position. Its network input is
\begin{equation}
    Z_i^{\mathrm{NTB}}=(\log(S_i/K),\tau_i,\sqrt{\tau_i},\Delta_i^{BS}).
\end{equation}
The network outputs a center shift $a_i^{\theta}$ and non-negative half-width $w_i^{\theta}$ around the Black--Scholes delta, so $z_i^{\theta}=\Delta_i^{BS}+a_i^{\theta}$ and
\begin{equation}
    q_i^{\mathrm{NTB}}=
    \min\{\max(q_{i-1}^{\mathrm{NTB}},z_i^{\theta}-w_i^{\theta}),
    z_i^{\theta}+w_i^{\theta}\}.
    \label{eq:learned-band}
\end{equation}
In implementation, $a_i^{\theta}=0.50\tanh(\cdot)$, $w_i^{\theta}=0.005+0.50\operatorname{sigmoid}(\cdot)$, and the center is clipped to $[-2,2]$ before projection. The carried position does not enter the NTB network input.

Fourth, the neural hedge can also be made sparse after training. Without a threshold, the raw policy trades on nearly every rebalancing date. The deep-hedging literature has also introduced explicit no-transaction filters. \citet{FrancoisGauthierGodin2025} jointly optimize a rebalancing threshold with the neural policy, whereas \citet{GongVentreOHara2021} filter trading dates using a pre-specified threshold on daily price changes. Here, by contrast, the network weights remain fixed and the execution threshold is selected afterward on a separate validation sample. 

Here we apply a calibrated threshold to the output of a trained neural hedger. At each date the proposal is computed using the previously executed sparse position,
\begin{equation}
    Z_i^h=(\log(S_i/K),\tau_i,\sqrt{\tau_i},q_{i-1}^{\theta,h}),
    \qquad \widetilde q_i^{\theta,h}=\Pi_{\theta}(Z_i^h).
\end{equation}
For $i=0,\ldots,N-1$, with $q_{-1}^{\theta,h}=0$,
\begin{equation}
    q_i^{\theta,h}=\begin{cases}
    \widetilde q_i^{\theta,h},&|\widetilde q_i^{\theta,h}-q_{i-1}^{\theta,h}|>h,\\
    q_{i-1}^{\theta,h},&|\widetilde q_i^{\theta,h}-q_{i-1}^{\theta,h}|\leq h.
    \end{cases}
    \label{eq:sparse-execution}
\end{equation}
The threshold is applied from inception and only proposed changes above it are executed. Appendix~\ref{app:execution} separately compares nearest-boundary projection with trigger-to-target execution for the classical bands.
 
\subsection{Tracking risk and robust HVA}
\label{sec:robusthva}
With the policy defined, this and the following subsections construct the post-training evaluation applied to each policy. Given policy $\pi$, let $P_0$ be the option price. The terminal tracking error and tracking risk are
\begin{align}
    E_T^{\pi}&=P_0+G_T^{\pi}-D_N(S_N-K)^+,\\
    R^{\pi}&=\CVaR_{95}[-E_T^{\pi}].
    \label{eq:tracking-risk}
\end{align}
Define $\Delta q_i^{\pi}=q_i^{\pi}-q_{i-1}^{\pi}$ and
\begin{equation}
    X_i^{\pi}=S_i|\Delta q_i^{\pi}|,
    \quad i=0,\ldots,N-1,
    \qquad
    X_N^{\pi}=S_N|q_{N-1}^{\pi}|.
\end{equation}
The pathwise hedging-friction cost is
\begin{equation}
    C^{\pi}=\sum_{i=0}^{N}D_iM_i
    \left(sX_i^{\pi}+\kappa(X_i^{\pi})^2\right).
    \label{eq:all-in-hva}
\end{equation}
Here $D_i$ is the discount factor, $s$ is the half-spread, $\kappa$ is the impact coefficient, and $M_i$ is the liquidity multiplier of Section~\ref{sec:model}. The sum includes the opening trade and Appendix~\ref{app:accounting} reports the continuation-cost convention which excludes the opening trade.
 
The relative-entropy (KL) upper expectation is a robust risk-measurement device for model misspecification \citep{GlassermanXu2014}. For each simulation path block $\mathcal S=\{\omega_n\}_{n=1}^{N_{\mathrm{sim}}}$, define the uniform empirical path law
\begin{equation}
    \widehat P_{\mathcal S}
    =\frac{1}{N_{\mathrm{sim}}}\sum_{n=1}^{N_{\mathrm{sim}}}\delta_{\omega_n},
    \label{eq:empirical-law}
\end{equation}
where $\delta_\omega$ denotes the Dirac measure at $\omega$ and $\widehat P_{\mathcal S}$ places equal mass $1/N_{\mathrm{sim}}$ on each simulated path. Following the robust-HVA construction of \citet{Sakuma2026RobustHVA}
\begin{equation}
    \RHVA^{\pi}(\varepsilon)=
    \sup_{Q:\KL(Q\|\widehat P_{\mathcal S})\leq\varepsilon}\E_Q[C^{\pi}]
    =\inf_{\eta>0}\left\{
    \eta\varepsilon+\eta\log\left(
    \frac{1}{N_{\mathrm{sim}}}\sum_{n=1}^{N_{\mathrm{sim}}}
    e^{C_n^{\pi}/\eta}\right)\right\}.
    \label{eq:robust-hva}
\end{equation}
The hedge path also creates two balance-sheet costs in the spirit of FVA and MVA: a funding charge on negative hedge cash balances and a margin charge on short-horizon hedged losses.
 
\subsection{Robust funding add-on}
\label{sec:robustfunding}
 
Following the funding and collateral valuation logic in \citet{Piterbarg2010} and \citet{PallaviciniPeriniBrigo2011}, negative hedge cash balances require funding. The post-trade hedge cash balance is
\begin{equation}
    F_0^{\pi}=P_0-q_0^{\pi}S_0
    -M_0\left(sX_0^{\pi}+\kappa(X_0^{\pi})^2\right),
\end{equation}
and, for $i=1,\ldots,N-1$,
\begin{equation}
    F_i^{\pi}=F_{i-1}^{\pi}-\Delta q_i^{\pi}S_i
    -M_i\left(sX_i^{\pi}+\kappa(X_i^{\pi})^2\right).
\end{equation}
Cash does not accrue at the risk-free rate and positive balances receive no credit. Only negative balances incur the annualized funding spread $s_f$ and the pathwise funding cost is
\begin{equation}
    C_{\mathrm{fund}}^{\pi}
    =\sum_{i=0}^{N-1}D_i s_f(F_i^{\pi})^-\Delta t,
    \label{eq:funding-cost}
\end{equation}
which represents a simplified policy-dependent funding add-on. The robust funding add-on is
\begin{equation}
    \RFA^{\pi}(\varepsilon)=
    \sup_{Q:\KL(Q\|\widehat P_{\mathcal S})\le\varepsilon}\E_Q[C_{\mathrm{fund}}^{\pi}].
    \label{eq:robust-funding}
\end{equation}
 
\subsection{Robust margin add-on}
\label{sec:robustmargin}
 
We follow \citet{GreenKenyon2015MVA}, in which initial margin creates a funding cost that depends on the expected margin profile. Let $H$ be the margin look-ahead horizon and $V_i$ a Black--Scholes option mark. For evaluation path $n$ and date $i$, the hedged mark-to-market loss over the horizon $H$ discounted to time $0$ is 
\begin{equation}
    \ell_{n,i:H}^{\pi}
    =(D_{i+H}V_{n,i+H}-D_iV_{n,i})
    -q_{n,i}^{\pi}(D_{i+H}S_{n,i+H}-D_iS_{n,i}).
\end{equation}
Let $\alpha\in(0,1)$ be the margin confidence level. With $k=\lceil(1-\alpha)N_{\mathrm{sim}}\rceil$ and $\mathcal I_i^{\pi}$ denoting the $k$ paths with the largest values of $\ell_{n,i:H}^{\pi}$, the time-$i$ margin profile is
\begin{equation}
    \widehat{IM}_i^{\pi}
    =\frac{1}{k}\sum_{n\in\mathcal I_i^{\pi}}(\ell_{n,i:H}^{\pi})^+.
    \label{eq:empirical-im}
\end{equation}
But $\widehat{IM}_i^{\pi}$ is already an average over paths and cannot be reweighted. Instead we need a pathwise allocation whose empirical average reproduces this margin profile. We therefore define
\begin{equation}
    C_{\mathrm{margin},n}^{\pi}
    =\sum_{i=0}^{N-H}s_m\frac{N_{\mathrm{sim}}}{k}
    (\ell_{n,i:H}^{\pi})^+\mathbf 1\{n\in\mathcal I_i^{\pi}\}\Delta t,
    \label{eq:margin-cost}
\end{equation}
where $s_m$ is the margin funding spread and the indicator $\mathbf 1\{n\in\mathcal I_i^{\pi}\}$ assigns the cost only to the $k$ paths that define the margin profile at date $i$. Averaging over all $N_{\mathrm{sim}}$ paths recovers the margin profile exactly, $N_{\mathrm{sim}}^{-1}\sum_n C_{\mathrm{margin},n}^{\pi}=\sum_{i=0}^{N-H}s_m\widehat{IM}_i^{\pi}\Delta t$. The sets $\mathcal I_i^{\pi}$ are formed under the reference empirical distribution and held fixed when $Q$ reweights the paths. With $C_{\mathrm{margin}}^{\pi}$ denoting the resulting pathwise random variable, the robust margin add-on is
\begin{equation}
    \RMA^{\pi}(\varepsilon)=
    \sup_{Q:\KL(Q\|\widehat P_{\mathcal S})\le\varepsilon}\E_Q[C_{\mathrm{margin}}^{\pi}].
    \label{eq:robust-margin}
\end{equation}
 
\subsection{Risk-constrained ranking}
\label{sec:riskrankings}
 
The ranking rule below evaluates HVA, funding, and margin costs under a single common-stress distribution. Define the total path cost
\begin{equation}
    A^{\pi}=C^{\pi}+C_{\mathrm{fund}}^{\pi}+C_{\mathrm{margin}}^{\pi}
\end{equation}
and the robust total adjustment
\begin{equation}
    \RAdj^{\pi}(\varepsilon)=
    \sup_{Q:\KL(Q\|\widehat P_{\mathcal S})\leq\varepsilon}\E_Q[A^{\pi}].
    \label{eq:robust-total}
\end{equation}
If $Q^{\star,\pi}$ solves \eqref{eq:robust-total}, we have
\begin{equation}
    \RAdj^{\pi}=
    \E_{Q^{\star,\pi}}[C^{\pi}]
    +\E_{Q^{\star,\pi}}[C_{\mathrm{fund}}^{\pi}]
    +\E_{Q^{\star,\pi}}[C_{\mathrm{margin}}^{\pi}].
    \label{eq:common-stress-decomposition}
\end{equation}
These common-stress components are distinct from the separately optimized robust quantities in equations~\eqref{eq:robust-hva}, \eqref{eq:robust-funding}, and~\eqref{eq:robust-margin} because each standalone supremum may imply a different worst-case path distribution. All policy rankings and reported reserve decompositions in this paper use the single optimizer $Q^{\star,\pi}$ of the total adjustment. Its empirical weights are
\begin{equation}
    w_n^{\star,\pi}
    =\frac{\exp(A_n^{\pi}/\eta^{\star,\pi})}
    {\sum_{m=1}^{N_{\mathrm{sim}}}\exp(A_m^{\pi}/\eta^{\star,\pi})},
    \label{eq:common-stress-weights}
\end{equation}
where $\eta^{\star,\pi}$ solves the dual problem for $A^{\pi}$.
We minimize the robust total adjustment subject to a tracking-risk budget:
\begin{equation}
    \min_{\pi}\ \RAdj^{\pi}
    \quad\text{subject to}\quad
    R^{\pi}\le c\,R^{\mathrm{ref}},
    \label{eq:riskcap}
\end{equation}
where $R^{\pi}=\CVaR_{95}[-E_T^{\pi}]$ and the multiplier $c$ is the trading desk's tracking-risk budget relative to a reference policy. The sum $\RHVA^{\pi}+\RFA^{\pi}+\RMA^{\pi}$ weakly exceeds $\RAdj^{\pi}$ and can overstate joint risk because the componentwise worst cases need not occur under the same path distribution. $\RAdj^{\pi}$ instead stresses the combined cost $A^{\pi}$ under one shared worst-case measure, and the same policy-specific tilt applies to the total and all three components. These are path-block-level empirical quantities and the KL reweighting is applied separately within each simulated path block.

\section{Numerical design}
\label{sec:design}
 
We use an at-the-money European call with $S_0=K=1$, maturity $T=1$, and $r=0.02$. Table~\ref{tab:markets} gives the three market environments. Middle and Low Liquidity add jump risk, higher volatility, more persistent stressed liquidity, wider spreads, and quadratic impact. For both fixed-width and gamma-wide rules, the base width $b_0$ is selected from
\begin{equation}
    b_0\in\{0.05,0.075,0.10,0.125,0.15,0.175,0.20\}.
    \label{eq:classical-grid}
\end{equation}
The sparse-execution threshold $h$ of \eqref{eq:sparse-execution} is selected from
\begin{equation}
    h\in\{0.05,0.075,0.10,0.125\}.
    \label{eq:sparse-grid}
\end{equation}
Applying these grids to the admissible policy families, together with the policy and learned NTB, yields 20 candidate specifications in High Liquidity and 27 in each jump environment. A classical specification contains one trading rule. A learned specification contains the three pre-specified independently trained checkpoints and is evaluated by averaging their metrics within each market-path seed.
 
\input{tables/table_markets.tex}

The learned policies are implemented in PyTorch. They follow \citet{Imaki2021PFHedge} but do not implement their PFHedge package \citep{PFHedge2024}. The direct and NTB networks have three hidden layers with 64 SiLU units per layer. The direct network bounds its proposed hedge at $|\widetilde q_i|\leq2$. The NTB network clips its center to $[-2,2]$. Each network was trained for 250 epochs with 16,384 paths per epoch and 64 trading dates. Optimization uses AdamW with learning rate $10^{-3}$, weight decay $10^{-5}$, and gradient clipping at 10. The coefficients on spread, impact, turnover, and soft trade count in \eqref{eq:training-objective} are 1, 1, 0.01, and $10^{-5}$.
 
The corresponding candidate specifications compete within each market and risk budget. Selecting the best candidate on the same simulated paths could overstate the performance and reward whichever candidate best fits the particular sample instead of the candidate which performs best in general. We avoid this by splitting the simulated paths into two disjoint sets. A validation set is used to pick a specification and a test set is used to report that specification's performance.

The frozen policies are evaluated at 252 dates and each market has five validation path blocks $\mathcal V$ and 20 out-of-validation test path blocks $\mathcal T$ with 5,000 paths per path block. The three training seeds are disjoint from these evaluation seeds. For candidate specification $g$, let $J_g=1$ for a classical rule and $J_g=3$ for a learned rule, and define its path-block-level metrics by
\begin{equation}
    \overline{\RAdj}_{\mathcal S}^{g}
    =\frac{1}{J_g}\sum_{j=1}^{J_g}\RAdj_{\mathcal S}^{\pi_{g,j}},
    \qquad
    \overline R_{\mathcal S}^{g}
    =\frac{1}{J_g}\sum_{j=1}^{J_g}R_{\mathcal S}^{\pi_{g,j}}.
    \label{eq:checkpoint-aggregation}
\end{equation}
Let $a_m$ denote the reference policy, defined market by market as the Black--Scholes $b=0.10$ band when $m$ is High Liquidity and the Merton $b=0.10$ band when $m$ is Middle or Low Liquidity. For budget multiple $c$, the five validation path blocks $\mathcal V$ select
\begin{equation}
    \widehat g_c=
    \arg\min_g\widehat{\overline{\RAdj}}_{\mathcal V}^{g}
    \quad\text{subject to}\quad
    \widehat{\overline R}_{\mathcal V}^{g}\leq
    c\widehat{\overline R}_{\mathcal V}^{a_m}.
    \label{eq:validation-selection}
\end{equation}
We evaluate the selected specification $\widehat g_c$ on the 20 disjoint test path blocks $\mathcal T$. A reported selection passes only if the same mean tracking-CVaR inequality also holds on $\mathcal T$. For a learned specification, this estimates average performance across the three fixed training runs.
 
$P_0$ is the analytic Merton value and is constant across seeds. The funding spread is $s_f=1\%$ per year. The margin add-on calculation uses $s_m=0.5\%$, $\alpha=99\%$, and a ten-step horizon. The numerical KL radii are the $b=0.10$ entries of the fixed benchmark-stress map at $\rho_0^G=0.4$ in \citet{Sakuma2026RobustHVA}, where $\rho_0^G$ is a Gaussian rank-correlation parameter for demand and liquidity. They are treated as external market-level stress inputs and held fixed across all candidate specifications. We calculate gamma intensity on an independent sample as in Table~\ref{tab:calibration}.
 
\input{tables/table_calibration.tex}

\section{Results}
\label{sec:results}
 
Section~\ref{sec:frontier} compares total adjustment and tracking-loss CVaR across the full classical-width grids and the reported learned specifications on the 20 test seeds. Section~\ref{sec:risk-selection} reports which candidate specification the validation sample selects and whether that selection still meets the risk budget on the test seeds. Section~\ref{sec:uncertainty} computes a confidence interval for the Low Liquidity reserve reduction from sparse execution and checks whether the reduction is stable across different training and validation seeds, with details in the appendix.
 
\subsection{Cost versus risk across all policies}
\label{sec:frontier}
This section reports the cost--risk frontier on the 20 test path blocks 
$\mathcal T$ (Section~\ref{sec:design}) for the candidate policy families before a risk-budget constraint is applied. Consider first the extreme case of no hedging. The no-hedge rule has little transaction-related reserve, but its tracking-loss CVaR is approximately 10.6, 8.5, and 7.0 times the reference policy's tracking-loss CVaR in High, Middle, and Low Liquidity. Ranking policies by reserve cost alone would favor this rule, even though it leaves the desk almost entirely exposed to price risk. Considering a tracking-risk constraint is necessary as discussed in Section~\ref{sec:risk-selection}.

Table~\ref{tab:core-results} reports test results for the candidate families. The three markets tell different stories. In High Liquidity, tuning the classical band already captures most of the achievable reserve reduction. The Black--Scholes $b=0.10$ reference policy has total adjustment 0.005243 and tracking CVaR 0.0392, and gamma-wide $b_0=0.075$ only changes these by -0.1\% and -3.0\%; the evaluated sparse and NTB rules do not improve on this tuned classical frontier. Middle Liquidity looks similar. The Merton $b=0.10$ reference policy has total adjustment 0.01204 and tracking CVaR 0.0680. Gamma-wide $b_0=0.075$ changes these by -1.6\% and -2.9\%. The evaluated sparse policy is not the frontier choice once the classical width is tuned. Low Liquidity is the exception. The Merton $b=0.10$ reference policy has total adjustment 0.1473 and tracking CVaR 0.1336. Sparse 0.075 lowers total adjustment to 0.1411, a 4.2\% reduction, with a mean tracking-CVaR change of -2.2\%. The paired total difference is negative in all 20 test seeds, with a 95\% interval of $[-0.007066,-0.005290]$. Wider classical bands produce larger reserve reductions as the risk budget is relaxed.
 
\input{tables/table_core_results.tex}

\begin{figure}[H]
    \centering
    \includegraphics[width=0.99\textwidth]{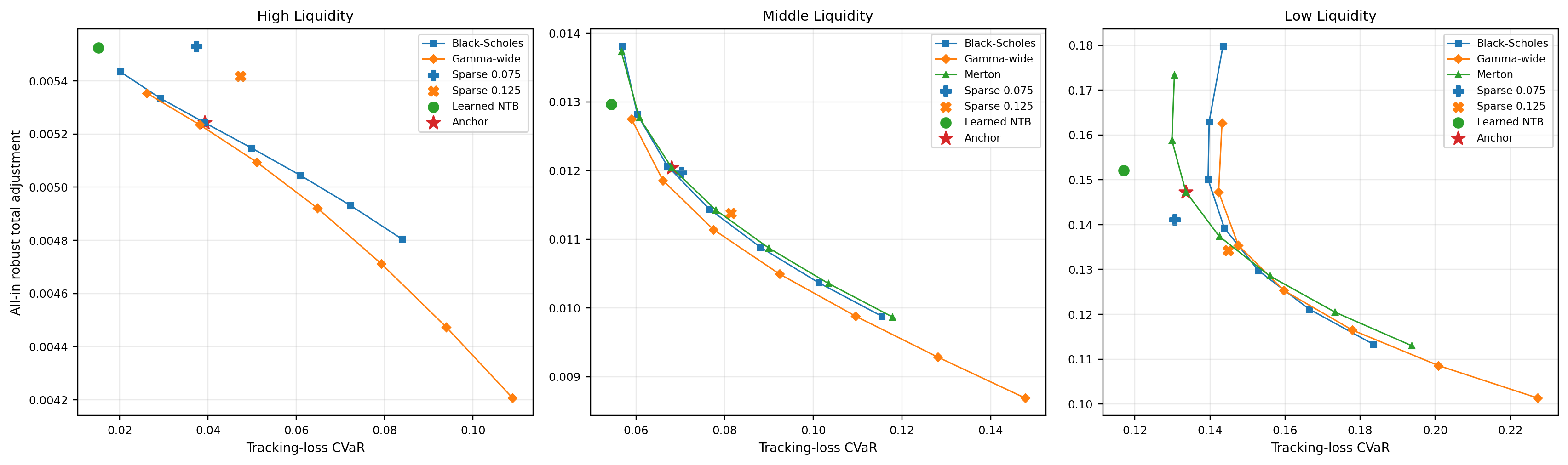}
    \caption{Out-of-validation full-lifecycle robust total adjustment against tracking-loss CVaR. Classical lines trace the fixed seven-point base-width grid.}
    \label{fig:frontier}
\end{figure}

Figure~\ref{fig:frontier} plots the classical-width curves and the learned-specification points. The figure highlights that sparse learned execution improves the strict Low Liquidity comparison, but once the classical width is tuned, a classical policy achieves the best relaxed-budget cost-risk trade-off in Low Liquidity.
 
\subsection{Specifications selected under the validation rule}
\label{sec:risk-selection}
 
This section applies the validation-based selection rule of Section~\ref{sec:design} at three risk budgets ($c=1$, $1.2$, $1.5$). For each budget, Table~\ref{tab:risk-selection} shows which candidate the validation rule selects and whether the choice still holds up on test data. At $c=1$, High and Middle Liquidity select gamma-wide $b_0=0.075$, while Low Liquidity selects sparse 0.075. All three selections satisfy the same mean tracking-CVaR budget on the test seeds. The learned-policy gain is confined to Low Liquidity, where total adjustment falls by 4.2\% relative to the Merton reference policy. At $c=1.2$, the selected specifications are gamma-wide $b_0=0.075$, $0.10$, and $0.125$ in High, Middle, and Low Liquidity; all pass the test-sample budget. At $c=1.5$, High Liquidity selects gamma-wide $b_0=0.10$ and passes, whereas the Middle Merton $b=0.175$ and Low gamma-wide $b_0=0.175$ selections exceed the test-sample budget slightly.
 
\input{tables/table_risk_selection.tex}

\begin{figure}[H]
    \centering
    \includegraphics[width=0.99\textwidth]{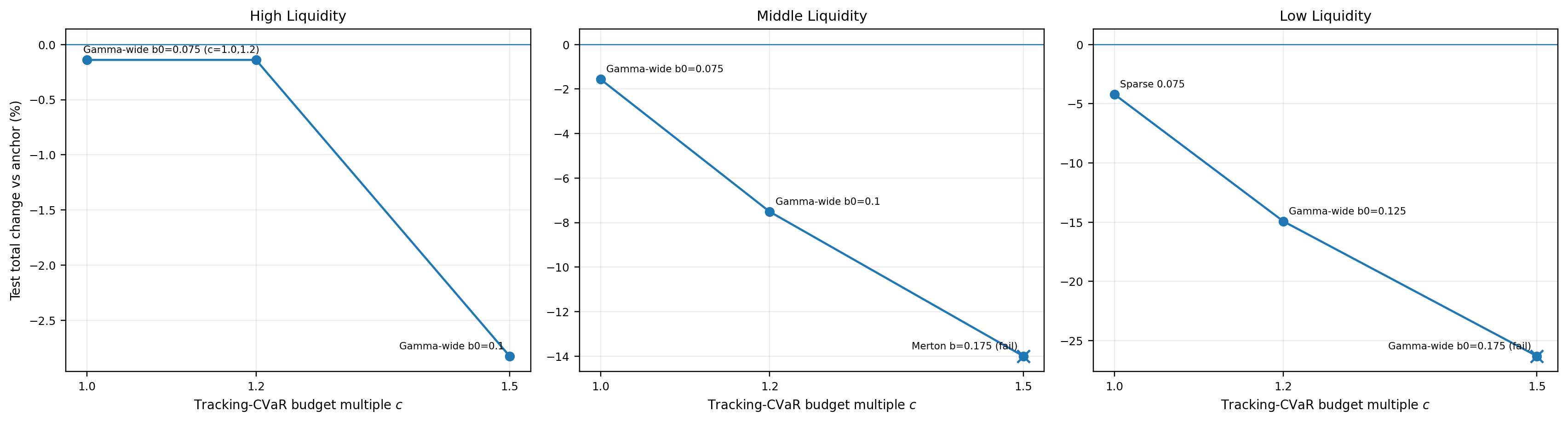}
    \caption{Out-of-validation total-adjustment change for each specification selected under the validation rule.}
    \label{fig:risk-selection}
\end{figure}

Figure~\ref{fig:risk-selection} plots, for each validation-selected specification, the resulting percentage change in total adjustment on the test seeds relative to the reference policy. A cross marks a specification whose test-sample risk exceeds the same risk budget it was selected under. Appendix~\ref{app:relaxed-grid} shows how each classical width in the grid performs at relaxed budgets.
 
\subsection{Uncertainty and stability relative to the reference policy}
\label{sec:uncertainty}
 
Section~\ref{sec:frontier} found that sparse execution lowers Low Liquidity cost by an average of 4.2\% across 20 test seeds, but an average alone could hide a difference that is smaller, zero, or reversed depending on the market-path seeds, trained network, or validation sample. Table~\ref{tab:paired-uncertainty} pairs each candidate outcome with the reference policy's outcome on the same simulated paths, removing path-to-path noise. The strict High and Middle gamma-wide comparisons have narrow intervals because only the band rule changes. In Low Liquidity, the sparse 0.075 total difference is negative in all 20 paired test seeds, and its tracking-CVaR difference is also negative in every seed, with a 95\% interval of $[-0.00497,-0.00101]$. These intervals measure Monte Carlo variation conditional on the fixed model, cost parameters, and trained checkpoints. Appendix~\ref{app:stability} checks its stability across seeds and networks.
 
\input{tables/table_paired_uncertainty.tex}

The stability checks in Appendix~\ref{app:stability} separate the fifteen seeds added after the earlier five-seed evaluation. On these newly added seeds sparse 0.075 reduces Low Liquidity total adjustment by 4.27\% and all fifteen differences are negative. Each of the three trained networks also lowers mean total adjustment relative to the reference policy with reductions from 3.50\% to 5.17\%. Finally, leaving out one validation seed at a time preserves the strict-budget selection in all five omissions for each environment.
 
Appendix~\ref{app:components} decomposes the reserve change. In Low Liquidity, sparse 0.075 changes stressed HVA, funding, and margin by -0.006180, -0.000022, and 0.000024. The reserve reduction therefore comes from transaction-cost HVA instead of a shift of cost across components that leaves the total unchanged.
 
\section{Discussion}
 
This paper evaluates deep-hedging policies with a valuation-adjustment layer and compares classical and learned candidate specifications under the same tracking-risk budget. Classical widths and learned sparse-execution thresholds are selected on the same validation sample, while neural weights and training hyperparameters remain fixed. The best policy depends on the market environment, and no single rule dominates across all three environments. In High and Middle Liquidity, a gamma-wide classical band remains competitive with the learned alternatives. In Low Liquidity, sparse learned execution still improves on the tuned classical band under a strict risk budget, though wider classical bands close most of that gap once the budget is relaxed.

Deep hedging and the valuation-adjustment layer play complementary roles. Deep hedging searches for state-dependent trading rules and no-transaction structure is added either inside the policy or at execution. A learned hedge is worth using only when it improves the cost-risk frontier after the tracking-risk constraint is imposed. The classical rule remains the better choice otherwise.

This High and Middle Liquidity result depends on tuning the classical band on the same footing as the learned methods. Sparse execution appeared to beat a fixed-width classical band at relaxed risk budgets in Low Liquidity. However, when we let the classical band's width be tuned on the same sample as the learned threshold, most of the advantage disappeared. The learned NTB is consistently risk-conservative but its moving center produces frequent trades.

In Low Liquidity, sparse execution keeps a modest advantage. But the Markov liquidity state never enters its training or its policy input and the gain more plausibly reflects a favorable interaction between the trained target and post-training filtering under severe frictions. 
 
Several limitations remain. The policies neither observe the Markov liquidity state nor train against its multiplier. The learned comparisons average three fixed checkpoints trained at 64 dates and transferred to 252-date evaluation; they do not identify a best usable checkpoint or the attainable frontier under fresh 252-date retraining. The training penalties also exclude the opening charge and Markov multiplier used in evaluation. The tracking-risk constraint is evaluated under the reference risk-neutral law; it is neither robustified over the KL neighborhood nor calibrated under a physical measure. The liquidity chain is simulated independently of the Brownian and jump innovations, so the experiment does not capture endogenous wrong-way dependence in which severe market moves and liquidity deterioration arrive together. The funding and margin add-ons are simplified. The KL radii are external fixed stress inputs taken from the prior $b=0.10$, $\rho_0^G=0.4$ map and are not recalibrated for the expanded total-cost definition or policy set.
 
Our framework can extend to a balance-sheet treatment. Our funding add-on can be tied to a treasury funding curve and a netting agreement. Our margin add-on can be tied to a desk IM model, for example a SIMM-style or ES-based rule. Collateral, netting, counterparty default, close-out, CVA capital, and KVA can be added as separate layers and the hedge policy changes exposure, collateral demand, CVA hedges, or capital usage. As an alternative to computing these layers separately, neural BSDE xVA solvers such as the deep xVA solver of \citet{GnoattoPicarelliReisinger2023} can compute the coupled system jointly. But such a solver remains subject to model risk \citep{BenezetCrepey2024}.
 
\section{Conclusion}
A valuation-adjustment layer changes the comparison of deep and classical hedging policies because transaction cost and tracking risk must be assessed jointly. By evaluating hedging valuation adjustment, funding, and margin together under one common worst-case stress, this framework gives a desk a single, internally consistent reserve for ranking classical and learned policies, rather than three reserves computed under different assumptions. Classical widths and sparse-execution thresholds are selected on the same validation sample, while neural weights and training hyperparameters remain fixed. No single candidate specification dominates across the three market environments. Classical bands are selected in the High- and Middle-Liquidity environments. Under the strict Low-Liquidity budget, the sparse learned specification reduces the reserve by a few percent beyond the best tuned classical band. Its reported value averages three fixed 64-date training runs transferred to 252-date evaluation.

\section*{Declaration of generative AI and AI-assisted technologies}
During the preparation of this manuscript, the author used generative AI and AI-assisted tools to assist with manuscript drafting, language refinement, code drafting and debugging, and discussion of the research. The author directed the project, revised and verified the manuscript and code, and takes full responsibility for the final manuscript.

\appendix
 
\section{Relaxed-budget descriptive grid points}
\label{app:relaxed-grid}
 
Section~\ref{sec:frontier} notes that wider classical bands reduce the reserve further once the risk budget is relaxed. Table~\ref{tab:relaxed-grid-examples} shows three classical grid points that illustrate the size of this effect. Each width belongs to the fixed seven-point candidate grid and passed the validation-sample budget.
 
\input{tables/table_relaxed_grid_examples.tex}

\section{Test-seed and training stability}
\label{app:stability}
 
The Low Liquidity result in Section~\ref{sec:frontier} may depend on the test seeds, trained network, and validation sample. Table~\ref{tab:test-seed-expansion} separates the earlier five test seeds, the fifteen newly added seeds, and the combined twenty. Table~\ref{tab:training-seed-stability} reports the Low Liquidity sparse result for each trained network. Table~\ref{tab:validation-loo} records leave-one-validation-seed-out selection stability.
 
\input{tables/table_test_seed_expansion.tex}

\input{tables/table_training_seed_stability.tex}

\input{tables/table_validation_loo.tex}

\section{Reserve component decomposition}
\label{app:components}
 
Section~\ref{sec:uncertainty} attributes the Low Liquidity reserve gain to transaction-cost HVA. Table~\ref{tab:component-changes} supports this by decomposing each policy's robust total under its own common worst-case path tilt. 
 
\input{tables/table_component_changes.tex}

\section{Initial-position sensitivity}
\label{app:initialization}
 
The main classical rule begins at $q_{-1}=0$ and projects a zero starting position to the nearest point of the initial band. This sensitivity check instead establishes the center target at inception and applies the same projected rule, as reported in Table~\ref{tab:initialization-sensitivity}. The base width is held fixed at the inherited $b_0=0.10$, so the comparison isolates initialization rather than width selection. Starting from the center target raises total adjustment in every environment. 
 
\input{tables/table_initialization_sensitivity.tex}

\section{Execution-rule sensitivity}
\label{app:execution}
 
The main results use nearest-boundary projection; this appendix examines how much that choice matters. Holding target initialization fixed, Table~\ref{tab:execution-sensitivity} compares nearest-boundary projection with trigger-to-target execution. The trigger rule trades to the center rather than the nearest boundary, and this increases turnover and reserve cost.
 
\input{tables/table_execution_sensitivity.tex}

\section{Continuation execution cost}
\label{app:accounting}
The main results count the opening trade's execution cost together with every rebalance and the terminal unwind. The continuation rule instead assumes the desk already holds the position it needs at date 0. Table~\ref{tab:accounting-sensitivity} shows that the difference is small in High Liquidity and material when spread and impact are large.
\input{tables/table_accounting_sensitivity.tex}

\section{Learned-band trading behavior}
\label{app:ntb}
 
The learned NTB trades whenever the carried position falls outside its updated state-dependent band. The band is reasonably wide, with a mean total width between 0.073 and 0.099 across the three markets. Yet because the band's center shifts at every date, a breach can occur without a large price move. Each breach needs a small trade to correct it, but these frequent small trades accumulate into higher turnover. Table~\ref{tab:ntb-diagnostics} reports band width, center movement, and the fraction of dates the previous hedge falls outside the band. Its moving center explains the low tracking risk and higher reserve cost. The post-training sparse-execution threshold and the NTB band both convert a neural hedge output into a sparse trading rule. However, sparse execution is the stronger cost-saving policy in Low Liquidity while the NTB band lowers tracking risk in Middle and Low Liquidity at the cost of trading more often.
 
\input{tables/table_ntb_diagnostics.tex}

\section{Monte Carlo variation}
\label{app:mc}
 
Table~\ref{tab:mc-variation} reports means and standard deviations across the 20 out-of-validation test seeds. Learned-specification values are averaged across the three fixed training seeds within each market-path seed. Because all policies are evaluated on common market-path seeds, Table~\ref{tab:paired-uncertainty} assesses policy differences using paired differences instead of marginal standard deviations.
 
\input{tables/table_mc_variation.tex}

\end{document}

%% file: tables/table_markets.tex
\begin{table}[H]
\centering
\scriptsize
\caption{Market and liquidity parameters. Jump sizes satisfy $\log J\sim N(\mu_J,\sigma_J^2)$.}
\label{tab:markets}
\resizebox{\textwidth}{!}{%
\begin{tabular}{lrrrrrrrrr}
\toprule
Environment & $\sigma$ & $\lambda_J$ & $\mu_J$ & $\sigma_J$ & Half-spread $s$ & Impact $\kappa$ & $m_s$ & $p_{NN}$ & $p_{SS}$ \\
\midrule
High Liquidity & 0.20 & 0.0 & -0.05 & 0.10 & 0.0001 & 0.0000 & 5 & 0.985 & 0.75 \\
Middle Liquidity & 0.25 & 0.5 & -0.06 & 0.12 & 0.0005 & 0.0015 & 10 & 0.975 & 0.90 \\
Low Liquidity & 0.30 & 2.0 & -0.10 & 0.18 & 0.0025 & 0.0080 & 20 & 0.960 & 0.95 \\
\bottomrule
\end{tabular}%
}
\end{table}

%% file: tables/table_calibration.tex
\begin{table}[H]
\centering
\small
\caption{Fixed evaluation inputs. $Q_{90}$ is estimated on 20,000 paths from an independent calibration sample.}
\label{tab:calibration}
\begin{tabular}{lrrr}
\toprule
Environment & $P_0$ & KL radius $\varepsilon$ & Gamma scale $Q_{90}$ \\
\midrule
High Liquidity & 0.089 & 0.006 & 0.041 \\
Middle Liquidity & 0.115 & 0.005 & 0.040 \\
Low Liquidity & 0.168 & 0.006 & 0.037 \\
\bottomrule
\end{tabular}
\end{table}

%% file: tables/table_core_results.tex
\begin{table}[H]
\centering
\footnotesize
\caption{Out-of-validation test comparison. Learned-specification entries average evaluation metrics across three independently trained networks within each market-path seed.}
\label{tab:core-results}
\begin{tabular}{llrrr}
\toprule
Environment & Policy & Robust total & Track CVaR & $\Delta$ total \\
\midrule
High Liquidity & Black--Scholes band $b=0.1$ & 0.005243 & 0.0392 & 0.0\% \\
High Liquidity & Gamma-wide $b_0=0.075$ & 0.005235 & 0.0380 & -0.1\% \\
High Liquidity & Gamma-wide $b_0=0.1$ & 0.005095 & 0.0510 & -2.8\% \\
High Liquidity & Sparse 0.075 & 0.005529 & 0.0374 & 5.5\% \\
High Liquidity & Learned NTB & 0.005524 & 0.0152 & 5.4\% \\
\midrule
Middle Liquidity & Merton band $b=0.1$ & 0.01204 & 0.0680 & 0.0\% \\
Middle Liquidity & Gamma-wide $b_0=0.075$ & 0.01186 & 0.0661 & -1.6\% \\
Middle Liquidity & Gamma-wide $b_0=0.1$ & 0.01114 & 0.0775 & -7.5\% \\
Middle Liquidity & Gamma-wide $b_0=0.125$ & 0.01049 & 0.0924 & -12.9\% \\
Middle Liquidity & Merton band $b=0.15$ & 0.01087 & 0.0900 & -9.7\% \\
Middle Liquidity & Merton band $b=0.175$ & 0.01036 & 0.1034 & -14.0\% \\
Middle Liquidity & Sparse 0.125 & 0.01138 & 0.0814 & -5.5\% \\
Middle Liquidity & Learned NTB & 0.01296 & 0.0544 & 7.6\% \\
\midrule
Low Liquidity & Merton band $b=0.1$ & 0.1473 & 0.1336 & 0.0\% \\
Low Liquidity & Sparse 0.075 & 0.1411 & 0.1306 & -4.2\% \\
Low Liquidity & Sparse 0.125 & 0.1342 & 0.1449 & -8.9\% \\
Low Liquidity & Gamma-wide $b_0=0.125$ & 0.1253 & 0.1597 & -14.9\% \\
Low Liquidity & Merton band $b=0.15$ & 0.1286 & 0.1559 & -12.7\% \\
Low Liquidity & Gamma-wide $b_0=0.175$ & 0.1086 & 0.2009 & -26.3\% \\
Low Liquidity & Merton band $b=0.2$ & 0.1130 & 0.1937 & -23.3\% \\
Low Liquidity & Learned NTB & 0.1521 & 0.1170 & 3.3\% \\
\bottomrule
\end{tabular}
\end{table}

%% file: tables/table_risk_selection.tex
\begin{table}[H]
\centering
\footnotesize
\caption{Validation-selected candidate specifications and out-of-validation test performance. Pass means the selected specification satisfies the same mean tracking-CVaR budget on the test seeds.}
\label{tab:risk-selection}
\begin{tabular}{lr l rrrc}
\toprule
Environment & $c$ & Selected specification & Test total & CVaR ratio & $\Delta$ total & Pass \\
\midrule
High Liquidity & 1.0 & Gamma-wide $b_0=0.075$ & 0.005235 & 0.970 & -0.1\% & Yes \\
High Liquidity & 1.2 & Gamma-wide $b_0=0.075$ & 0.005235 & 0.970 & -0.1\% & Yes \\
High Liquidity & 1.5 & Gamma-wide $b_0=0.1$ & 0.005095 & 1.300 & -2.8\% & Yes \\
\midrule
Middle Liquidity & 1.0 & Gamma-wide $b_0=0.075$ & 0.01186 & 0.971 & -1.6\% & Yes \\
Middle Liquidity & 1.2 & Gamma-wide $b_0=0.1$ & 0.01114 & 1.139 & -7.5\% & Yes \\
Middle Liquidity & 1.5 & Merton band $b=0.175$ & 0.01036 & 1.519 & -14.0\% & No \\
\midrule
Low Liquidity & 1.0 & Sparse 0.075 & 0.1411 & 0.978 & -4.2\% & Yes \\
Low Liquidity & 1.2 & Gamma-wide $b_0=0.125$ & 0.1253 & 1.195 & -14.9\% & Yes \\
Low Liquidity & 1.5 & Gamma-wide $b_0=0.175$ & 0.1086 & 1.504 & -26.3\% & No \\
\bottomrule
\end{tabular}
\end{table}

%% file: tables/table_paired_uncertainty.tex
\begin{table}[H]
\centering
\scriptsize
\caption{Paired out-of-validation test differences relative to the reference policy. Intervals are two-sided 95\% Student-$t$ intervals across common market-path seeds.}
\label{tab:paired-uncertainty}
\resizebox{\textwidth}{!}{%
\begin{tabular}{llrrrrr}
\toprule
Environment & Policy & Mean $\Delta$ total & 95\% CI & Mean $\Delta$ CVaR & 95\% CI & Total $<0$ \\
\midrule
High Liquidity & Gamma-wide $b_0=0.075$ & -0.000007 & [-0.000008, -0.000007] & -0.00118 & [-0.00139, -0.00098] & 20/20 \\
Middle Liquidity & Gamma-wide $b_0=0.075$ & -0.000188 & [-0.000195, -0.000182] & -0.00195 & [-0.00231, -0.00159] & 20/20 \\
Low Liquidity & Sparse 0.075 & -0.006178 & [-0.007066, -0.005290] & -0.00299 & [-0.00497, -0.00101] & 20/20 \\
Middle Liquidity & Gamma-wide $b_0=0.125$ & -0.001550 & [-0.001557, -0.001542] & 0.02434 & [0.02373, 0.02494] & 20/20 \\
Low Liquidity & Merton band $b=0.15$ & -0.01874 & [-0.01892, -0.01856] & 0.02234 & [0.02126, 0.02343] & 20/20 \\
Low Liquidity & Merton band $b=0.2$ & -0.03430 & [-0.03465, -0.03395] & 0.06013 & [0.05789, 0.06238] & 20/20 \\
\bottomrule
\end{tabular}%
}
\end{table}

%% file: tables/table_relaxed_grid_examples.tex
\begin{table}[H]
\centering
\scriptsize
\caption{Post-selection descriptive classical grid points at relaxed budgets.  }
\label{tab:relaxed-grid-examples}
\resizebox{\textwidth}{!}{%
\begin{tabular}{lrllrrrrrr}
\toprule
Environment & $c$ & Policy & Val. ratio & Val. total & Test ratio & Test total & $\Delta$ total & Val. pass & Test pass \\
\midrule
Middle Liquidity & 1.5 & Gamma-wide $b_0=0.125$ & 1.331 & 0.01040 & 1.358 & 0.01049 & -12.9\% & Yes & Yes \\
Low Liquidity & 1.2 & Merton band $b=0.15$ & 1.159 & 0.1288 & 1.167 & 0.1286 & -12.7\% & Yes & Yes \\
Low Liquidity & 1.5 & Merton band $b=0.2$ & 1.443 & 0.1132 & 1.450 & 0.1130 & -23.3\% & Yes & Yes \\
\bottomrule
\end{tabular}%
}
\end{table}

%% file: tables/table_test_seed_expansion.tex
\begin{table}[H]
\centering
\scriptsize
\caption{Low Liquidity sensitivity to test-seed expansion for Sparse 0.075 relative to the Merton $b=0.10$ reference policy. The fifteen seeds 1011--1025 were added after the earlier five-seed evaluation.}
\label{tab:test-seed-expansion}
\resizebox{\textwidth}{!}{%
\begin{tabular}{lrrrrrrr}
\toprule
Test-seed subset & $n$ & Anchor total & Sparse total & $\Delta$ total & Mean paired $\Delta$ & 95\% CI & $\Delta<0$ \\
\midrule
Earlier five (1006--1010) & 5 & 0.1494 & 0.1435 & -3.97\% & -0.005938 & [-0.008428, -0.003449] & 5/5 \\
Newly added fifteen (1011--1025) & 15 & 0.1466 & 0.1403 & -4.27\% & -0.006258 & [-0.007325, -0.005191] & 15/15 \\
Combined twenty (1006--1025) & 20 & 0.1473 & 0.1411 & -4.19\% & -0.006178 & [-0.007066, -0.005290] & 20/20 \\
\bottomrule
\end{tabular}%
}
\end{table}

%% file: tables/table_training_seed_stability.tex
\begin{table}[H]
\centering
\small
\caption{Low Liquidity Sparse 0.075 by trained network on the twenty out-of-validation test seeds. Rows are separate trained networks used on their own; the main text averages their evaluation metrics and does not form a combined position across networks.}
\label{tab:training-seed-stability}
\begin{tabular}{rrrrrr}
\toprule
Training seed & Robust total & $\Delta$ total & Track CVaR & $\Delta$ CVaR & Total $<0$ \\
\midrule
20260619 & 0.1416 & -3.91\% & 0.1303 & -2.45\% & 20/20 \\
20260620 & 0.1421 & -3.50\% & 0.1302 & -2.51\% & 19/20 \\
20260621 & 0.1397 & -5.17\% & 0.1312 & -1.76\% & 20/20 \\
\bottomrule
\end{tabular}
\end{table}

%% file: tables/table_validation_loo.tex
\begin{table}[H]
\centering
\small
\caption{Leave-one-validation-seed-out stability of the strict-budget specification selection.}
\label{tab:validation-loo}
\begin{tabular}{llr}
\toprule
Environment & Full validation selection & Same selection \\
\midrule
High Liquidity & Gamma-wide $b_0=0.075$ & 5/5 \\
Middle Liquidity & Gamma-wide $b_0=0.075$ & 5/5 \\
Low Liquidity & Sparse 0.075 & 5/5 \\
\bottomrule
\end{tabular}
\end{table}

%% file: tables/table_component_changes.tex
\begin{table}[H]
\centering
\scriptsize
\caption{Out-of-validation test component changes relative to the reference policy. Components use the common policy-specific KL tilt underlying each total.}
\label{tab:component-changes}
\resizebox{\textwidth}{!}{%
\begin{tabular}{llrrrrrr}
\toprule
Environment & Policy & $\Delta$HVA & $\Delta$Funding & $\Delta$Margin & $\Delta$Total & $\Delta$Total (\%) & $\Delta$CVaR (\%) \\
\midrule
High Liquidity & Gamma-wide $b_0=0.075$ & -0.000021 & 0.000005 & 0.000009 & -0.000007 & -0.1\% & -3.0\% \\
High Liquidity & Gamma-wide $b_0=0.1$ & -0.000041 & -0.000131 & 0.000024 & -0.000148 & -2.8\% & 30.0\% \\
High Liquidity & Sparse 0.075 & 0.000013 & 0.000266 & 0.000008 & 0.000286 & 5.5\% & -4.7\% \\
\midrule
Middle Liquidity & Gamma-wide $b_0=0.075$ & -0.000182 & -0.000013 & 0.000007 & -0.000188 & -1.6\% & -2.9\% \\
Middle Liquidity & Gamma-wide $b_0=0.1$ & -0.000781 & -0.000144 & 0.000021 & -0.000905 & -7.5\% & 13.9\% \\
Middle Liquidity & Gamma-wide $b_0=0.125$ & -0.001274 & -0.000313 & 0.000038 & -0.001550 & -12.9\% & 35.8\% \\
\midrule
Low Liquidity & Sparse 0.075 & -0.006180 & -0.000022 & 0.000024 & -0.006178 & -4.2\% & -2.2\% \\
Low Liquidity & Merton band $b=0.15$ & -0.01846 & -0.000288 & 0.000008 & -0.01874 & -12.7\% & 16.7\% \\
Low Liquidity & Merton band $b=0.2$ & -0.03372 & -0.000608 & 0.000030 & -0.03430 & -23.3\% & 45.0\% \\
\bottomrule
\end{tabular}%
}
\end{table}

%% file: tables/table_initialization_sensitivity.tex
\begin{table}[H]
\centering
\scriptsize
\caption{Initial-position sensitivity. The main rule projects $q_{-1}=0$ to the nearest band boundary; the sensitivity establishes the center target at inception.}
\label{tab:initialization-sensitivity}
\resizebox{\textwidth}{!}{%
\begin{tabular}{llrrrrrr}
\toprule
Environment & Band & Zero-start total & Target-start total & $\Delta$ total & Zero-start CVaR & Target-start CVaR & Opening cost: zero / target \\
\midrule
High Liquidity & Black--Scholes band $b=0.1$ & 0.005243 & 0.005304 & 1.2\% & 0.0392 & 0.0378 & 0.000059 / 0.000071 \\
High Liquidity & Gamma-wide $b_0=0.1$ & 0.005095 & 0.005289 & 3.8\% & 0.0510 & 0.0463 & 0.000051 / 0.000071 \\
\midrule
Middle Liquidity & Merton band $b=0.1$ & 0.01204 & 0.01265 & 5.0\% & 0.0680 & 0.0670 & 0.001673 / 0.002263 \\
Middle Liquidity & Gamma-wide $b_0=0.1$ & 0.01114 & 0.01215 & 9.1\% & 0.0775 & 0.0729 & 0.001326 / 0.002232 \\
\midrule
Low Liquidity & Merton band $b=0.1$ & 0.1473 & 0.1574 & 6.9\% & 0.1336 & 0.1320 & 0.03228 / 0.04320 \\
Low Liquidity & Gamma-wide $b_0=0.1$ & 0.1354 & 0.1502 & 11.0\% & 0.1476 & 0.1397 & 0.02326 / 0.03979 \\
\bottomrule
\end{tabular}%
}
\end{table}

%% file: tables/table_execution_sensitivity.tex
\begin{table}[H]
\centering
\small
\caption{Interior execution-rule sensitivity under target initialization. A projected rule trades to the nearest boundary and trigger-to-target jumps to the center.}
\label{tab:execution-sensitivity}
\resizebox{\textwidth}{!}{%
\begin{tabular}{llrrrrr}
\toprule
Environment & Target & Projected total & Trigger total & $\Delta$ total & Projected CVaR & Trigger CVaR \\
\midrule
High Liquidity & Black--Scholes & 0.005304 & 0.005467 & 3.1\% & 0.0378 & 0.0238 \\
\midrule
Middle Liquidity & Black--Scholes & 0.01267 & 0.01501 & 18.5\% & 0.0659 & 0.0600 \\
Middle Liquidity & Merton & 0.01265 & 0.01487 & 17.5\% & 0.0670 & 0.0598 \\
\midrule
Low Liquidity & Black--Scholes & 0.1596 & 0.2005 & 25.6\% & 0.1376 & 0.1475 \\
Low Liquidity & Merton & 0.1574 & 0.1908 & 21.2\% & 0.1320 & 0.1350 \\
\bottomrule
\end{tabular}%
}
\end{table}

%% file: tables/table_accounting_sensitivity.tex
\begin{table}[H]
\centering
\scriptsize
\caption{Full-lifecycle versus continuation execution cost. Continuation excludes the opening execution charge and its funding-account effect; other inputs are unchanged.}
\label{tab:accounting-sensitivity}
\resizebox{\textwidth}{!}{%
\begin{tabular}{llrrrrr}
\toprule
Environment & Policy & Full-lifecycle total & Continuation total & Full-lifecycle increment & Increment (\%) & Mean opening cost \\
\midrule
High Liquidity & Black--Scholes band $b=0.1$ & 0.005243 & 0.005183 & 0.000059 & 1.1\% & 0.000059 \\
High Liquidity & Gamma-wide $b_0=0.075$ & 0.005235 & 0.005179 & 0.000057 & 1.1\% & 0.000056 \\
High Liquidity & Gamma-wide $b_0=0.1$ & 0.005095 & 0.005043 & 0.000052 & 1.0\% & 0.000051 \\
\midrule
Middle Liquidity & Merton band $b=0.1$ & 0.01204 & 0.01033 & 0.001717 & 16.6\% & 0.001673 \\
Middle Liquidity & Gamma-wide $b_0=0.075$ & 0.01186 & 0.01029 & 0.001569 & 15.3\% & 0.001531 \\
Middle Liquidity & Gamma-wide $b_0=0.125$ & 0.01049 & 0.009336 & 0.001158 & 12.4\% & 0.001133 \\
\midrule
Low Liquidity & Merton band $b=0.1$ & 0.1473 & 0.1143 & 0.03299 & 28.9\% & 0.03228 \\
Low Liquidity & Sparse 0.075 & 0.1411 & 0.1070 & 0.03410 & 31.9\% & 0.03344 \\
Low Liquidity & Merton band $b=0.15$ & 0.1286 & 0.1006 & 0.02797 & 27.8\% & 0.02739 \\
Low Liquidity & Merton band $b=0.2$ & 0.1130 & 0.08966 & 0.02335 & 26.0\% & 0.02288 \\
\bottomrule
\end{tabular}%
}
\end{table}

%% file: tables/table_ntb_diagnostics.tex
\begin{table}[H]
\centering
\small
\caption{Learned no-transaction-band policy. }
\label{tab:ntb-diagnostics}
\resizebox{\textwidth}{!}{%
\begin{tabular}{lrrrrr}
\toprule
Environment & Mean total width & Center move & Previous hedge outside & Turnover & Trades \\
\midrule
High Liquidity & 0.073 & 0.019 & 35.2\% & 2.864 & 89.8 \\
Middle Liquidity & 0.099 & 0.018 & 27.4\% & 2.328 & 70.1 \\
Low Liquidity & 0.081 & 0.014 & 25.3\% & 2.065 & 64.9 \\
\bottomrule
\end{tabular}%
}
\end{table}

%% file: tables/table_mc_variation.tex
\begin{table}[H]
\centering
\small
\caption{Variation across the 20 out-of-validation test market-path seeds. }
\label{tab:mc-variation}
\resizebox{\textwidth}{!}{%
\begin{tabular}{llrrr}
\toprule
Environment & Policy & Robust total & Track CVaR & Opening cost \\
\midrule
High Liquidity & Black--Scholes band $b=0.1$ & 0.005243 (0.000021) & 0.0392 (0.0011) & 0.000059 (0.000001) \\
High Liquidity & Gamma-wide $b_0=0.075$ & 0.005235 (0.000022) & 0.0380 (0.0009) & 0.000056 (0.000001) \\
High Liquidity & Gamma-wide $b_0=0.1$ & 0.005095 (0.000020) & 0.0510 (0.0012) & 0.000051 (0.000001) \\
\midrule
Middle Liquidity & Merton band $b=0.1$ & 0.01204 (0.000119) & 0.0680 (0.0019) & 0.001673 (0.000037) \\
Middle Liquidity & Gamma-wide $b_0=0.075$ & 0.01186 (0.000129) & 0.0661 (0.0021) & 0.001531 (0.000034) \\
Middle Liquidity & Gamma-wide $b_0=0.125$ & 0.01049 (0.000116) & 0.0924 (0.0020) & 0.001133 (0.000025) \\
\midrule
Low Liquidity & Merton band $b=0.1$ & 0.1473 (0.003554) & 0.1336 (0.0036) & 0.03228 (0.000599) \\
Low Liquidity & Sparse 0.075 & 0.1411 (0.004167) & 0.1306 (0.0029) & 0.03344 (0.000621) \\
Low Liquidity & Merton band $b=0.15$ & 0.1286 (0.003179) & 0.1559 (0.0043) & 0.02739 (0.000509) \\
Low Liquidity & Merton band $b=0.2$ & 0.1130 (0.002821) & 0.1937 (0.0058) & 0.02288 (0.000425) \\
\bottomrule
\end{tabular}%
}
\end{table}